\documentclass[aps,prl,reprint,groupedaddress,showpacs]{revtex4-1}
\usepackage{amsmath}
\usepackage{bm}
\usepackage{graphicx}
\usepackage{longtable}
\RequirePackage{ifthen}

\usepackage[usenames,dvipsnames]{color}
\usepackage{ulem} 

\newcommand{\green}{\textcolor{OliveGreen}}

\newcommand{\me}[3]{\langle #1 \vert #2 \vert #3 \rangle}
\newcommand{\ket}[1]{\vert #1 \rangle}

\begin{document}

\title{Surface polarization and edge charges}


\author{Yuanjun Zhou}
\affiliation{Department of Physics and Astronomy, Rutgers University, Piscataway, New Jersey 08854, USA}
\author{Karin Rabe}
\affiliation{Department of Physics and Astronomy, Rutgers University, Piscataway, New Jersey 08854, USA}
\author{David Vanderbilt}
\affiliation{Department of Physics and Astronomy, Rutgers University, Piscataway, New Jersey 08854, USA}

\date{\today}

\begin{abstract}
The term ``surface polarization'' is introduced to describe
the in-plane polarization existing at the surface of an insulating
crystal when the in-plane surface inversion symmetry is broken.
Here, the surface polarization is formulated in terms of a Berry
phase, with the hybrid Wannier representation providing a natural
basis for study of this effect. Tight binding models are used to
demonstrate how the surface polarization reveals itself via the
accumulation of charges at the corners/edges for a two dimensional
rectangular lattice and for GaAs.
\end{abstract}

\pacs{77.22.Ej, 73.20.-r, 71.15.-m}


\maketitle

\def\scr{\scriptsize}
\def\draftversion{false}
\def\draftversion{true}
\ifthenelse{\equal{\draftversion}{true}}{
  \marginparwidth 2.7in
  \marginparsep 0.5in
  \newcounter{comm} 
  \def\commnext{\stepcounter{comm}}
  \def\commtext{{\bf\color{blue}[\arabic{comm}]}}
  \def\commmar{{\bf\color{blue}[\arabic{comm}]}}
  \def\dvm#1{\commnext\marginpar{\small DV\commmar: #1}\commtext}
  \def\yjm#1{\commnext\marginpar{\small YJZ\commmar: #1}\commtext}
  \def\krm#1{\commnext\marginpar{\small KMR\commmar: #1}\commtext}
  \def\sm#1{\commnext\marginpar{\scr\green{SAVED\commmar: #1}}\commtext}
  \def\tnewpage{\marginpar{\small Temporary newpage}\newpage}
}{
  \def\dvm#1{}
  \def\yjm#1{}
  \def\krm#1{}
  \def\sm#1{}
  \def\tnewpage{}
}


\newcommand{\bP}{\bm{P}}
\newcommand{\cP}{\mathcal{P}}
\newcommand{\bcP}{\bm{\cP}}
\newcommand{\cpp}{\mathcal{p}}
\newcommand{\nhat}{\hat{\mathbf{n}}}

\begin{figure}
\includegraphics[width=\columnwidth]{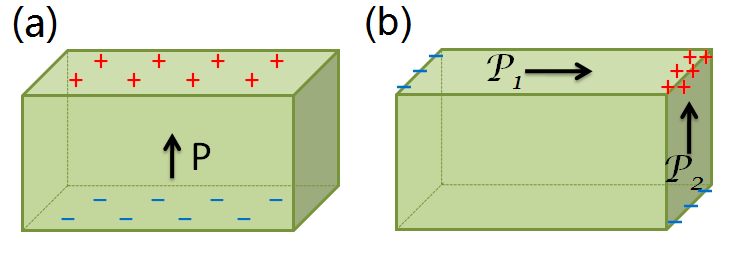}%
\caption{\label{definition}
Illustration of bulk and surface polarization effects.
The polarizations are denoted by black arrows, and
net positive and negative bounded charged are in red and blue,
respectively.
(a) Bulk polarization gives rise to surface charges $\sigma$.
(b) Surface polarization gives rise to edge charges $\lambda$.}
\end{figure}

For over two decades, it has been understood that the electric
polarization $\bf P$ of an insulating
crystal is a bulk quantity whose electronic contribution is
determined modulo $2e\bf R$/$\Omega$
(where $\bf R$ is a lattice vector and $\Omega$ is the unit cell volume)
by the Bloch functions through a Berry-phase expression,
or alternatively, in real space through the charge centers of
the Wannier
functions~\cite{polar-KSV-1993,polar-RMP-1994}. It was also shown that the macroscopic surface charge
of an insulating crystal is predicted by the standard
bound-charge expression $\sigma^{\rm surf}=\bf P\cdot\nhat$
(where $\nhat$ is the surface normal)~\cite{polar-VKS-1993},
as illustrated schematically in Fig.~\ref{definition}(a).

Here, we introduce and analyze a related quantity,
the \textit{``surface polarization,"}
defined as a 2-vector $\bcP$ lying in the plane of
an insulating surface of an insulating crystal.
By analogy with the bulk 3-vector $\bP$, it
has the property that when two facets meet, the linear
bound-charge density appearing on the shared edge is predicted to be
\begin{equation}
\lambda^{\textrm{edge}}=\bcP_1\cdot\nhat_1+ \bcP_2\cdot\nhat_2
\label{eq:edgecharge}
\end{equation}
where $\bcP_j$ is the surface polarization on facet $j$ and
$\nhat_j$ is a unit vector lying in the plane of the facet
and pointing toward (and normal to) the edge, as illustrated
in Fig.~\ref{definition}(b).

This surface polarization $\bcP$ is quite distinct from the
dipole per unit area \textit{normal} to the surface,
which has also been called ``surface polarization'' by
other authors~\cite{surfP-prb,surfacePNMat}. The latter is
always present regardless of the symmetry of the surface, and manifests itself
macroscopically through the surface work function.  In contrast,
our surface polarization $\bcP$ lies in-plane and is nonzero only
when the symmetry of the terminating surface supports a nonzero
in-plane vector, as for example on the (110) surface of GaAs.
It can also arise from a spontaneous symmetry-lowering surface
reconstruction, as observed recently at the
Pb$_{1-x}$Sn$_x$Se (110) surface~\cite{Science-TCI-surfaces}
and predicted for an ultrathin film of
SrCrO$_3$ on SrTiO$_3$ substrate (001) \cite{SCOferro}.
The surface polarization will be most evident when the
bulk $\bP$ vanishes, as will be the case for the systems
discussed below.

The purpose of this Letter is to extend the Berry-phase theory to
the case of surface polarization $\bcP$ as defined above.  To do
this, we introduce a formulation based on hybrid Wannier functions
(HWFs), which are Bloch-like parallel to the surface and Wannier-like in the surface-normal
direction~\cite{Resta-HWF,wu-prl06,soluyanov-prb11b,taherinejad-prb13}.
This allows for the use of Berry-phase methods parallel to
the surface while allowing a real-space identification of
the surface-specific contribution in the normal direction.
We illustrate the concept first for
a ``toy" 2D tightbinding (TB) model, demonstrating the method of calculating the
surface polarization.  We then consider an atomistic 3D model of
an ideal (110) surface of a generic III-V zincblende
semiconductor, using a TB model of GaAs to describe the electronic
structure. In both cases, we confirm that the surface polarization
correctly predicts corner and edge charges.


We first show how to express the surface polarization in terms of the Berry
phases of HWFs for a 2D insulating sample, which we take to lie in the
$(x,z)$ plane.
%
%
We take the ``surface'' (here really an edge)
to be normal to $\hat{z}$ and introduce HWFs $\ket{h_{lj}(k_x)}$,
where $l$ indexes unit-cell layers normal to the $z$ direction
and $j$ runs over occupied Wannier functions in a single unit cell.
For the bulk, the lattice is periodic in $z$ as well as $x$, and
the $\ket{h_{lj}(k_x)}$ and their centers $z^{\rm bulk}_{lj}(k_x)$
can be obtained using the 1D construction
procedure given in Ref.~\cite{Marzari-MLWF}.
To study the surface behavior we consider
a ribbon consisting of a finite number of unit cells along $\hat{z}$.
We then construct and diagonalize the matrix
$Z_{mn}=\me{\psi_m(k_x)}{z}{\psi_n(k_x)}$, whose eigenvectors yield
the HWFs
%
%
and whose
eigenvalues give the HWF centers $z_{lj}(k_x)$.  In practice these are
easily identified with the bulk $z^{\rm bulk}_{lj}(k_x)$ covering the range
of $l$ values that define the ribbon, with only modest shifts induced by
the presence of the surface, allowing a common labeling scheme for both.

If we were interested in computing the dipole moment \textit{normal} to the
surface, we could obtain this from an analysis of the $k_x$-averaged
$z$ positions $\bar z_{lj}$ of the HWFs,
where $a$ is the lattice constant along $x$.
However, our purpose here is different: we want to compute the polarization
\textit{parallel} to the surface.  For this, we compute the
Berry phase
\begin{equation}
\label{xpos}
 \gamma_{x,lj}=\int dk_x\,\me{h_{lj}}{i\frac{d}{dk_x}}{h_{lj}}.
\end{equation}
of each HWF ``band'' $(lj)$ as $k_x$ runs across the 1D BZ.
Doing the same for the bulk HWFs (these are independent of $l$)
and taking the difference, we obtain a set of Berry-phase shifts
$\Delta\gamma_{x,lj} \equiv \gamma_{x,lj}- \gamma_{x,j}^{\rm bulk}$
from which the electronic surface polarization can be determined via
\begin{equation}
\label{surfP}
\cP_x^{\rm elec}=-\frac{ea}{\Omega\pi}
  \sum_{lj=\textrm{center}}^{\textrm{top surf}}\Delta\gamma_{x,lj}
\end{equation}
where a factor of two has been included for spin degeneracy and
$\Omega$ is the edge repeat length $a$ in 2D.
%
%
Since $\Delta\gamma_{x,lj}$ decays exponentially into the bulk, 
the sum will converge within a few layers of the surface, but for definiteness
we sum to the center of the ribbon.
If the $z_{lj}$ values of some neighboring HWF bands overlap,
the procedure needs to be generalized by grouping the HWFs into
layers and using a multiband generalization to
assign contributions to each layer.

%
The generalization to a 3D crystal with surface normal to $z$ is
straightforward.  The HWFs are $\ket{h_{lj}(k_x,k_y)}$ with centers
$z_{lj}(k_x,k_y)$.  The surface polarization
$\cP_x^{\rm elec}$ is then obtained by computing Berry phases with respect
to $k_x$ as before, averaging over all $k_y$, and multiplying by the lattice
constant $a$ divided by the surface cell area $\Omega$. The other surface polarization $\cP_y$ is
given by the same formalism but with $x$ and $y$ reversed.

In the models considered in this paper, the surface polarization is purely electronic,
as the ions are held fixed in their bulk positions. More generally,
$\cP_x=\cP^{\rm ion}_x+\cP^{\rm elec}_x$ with the ionic contribution
given by
%
%
$\cP_x^{\rm ion}=\Omega^{-1}
\sum_{l\tau}Z_{\tau}\,(X_{l\tau}-X^{\rm bulk}_{l\tau})$,
where  $Z_\tau$ and $X_{l\tau}$ are the $x$ position and bare charge
of ion $\tau$ in cell $l$, and
$X^{\rm bulk}_\tau$ is the corresponding bulk position of the same atom.

\begin{figure}
 \includegraphics[width=\columnwidth]{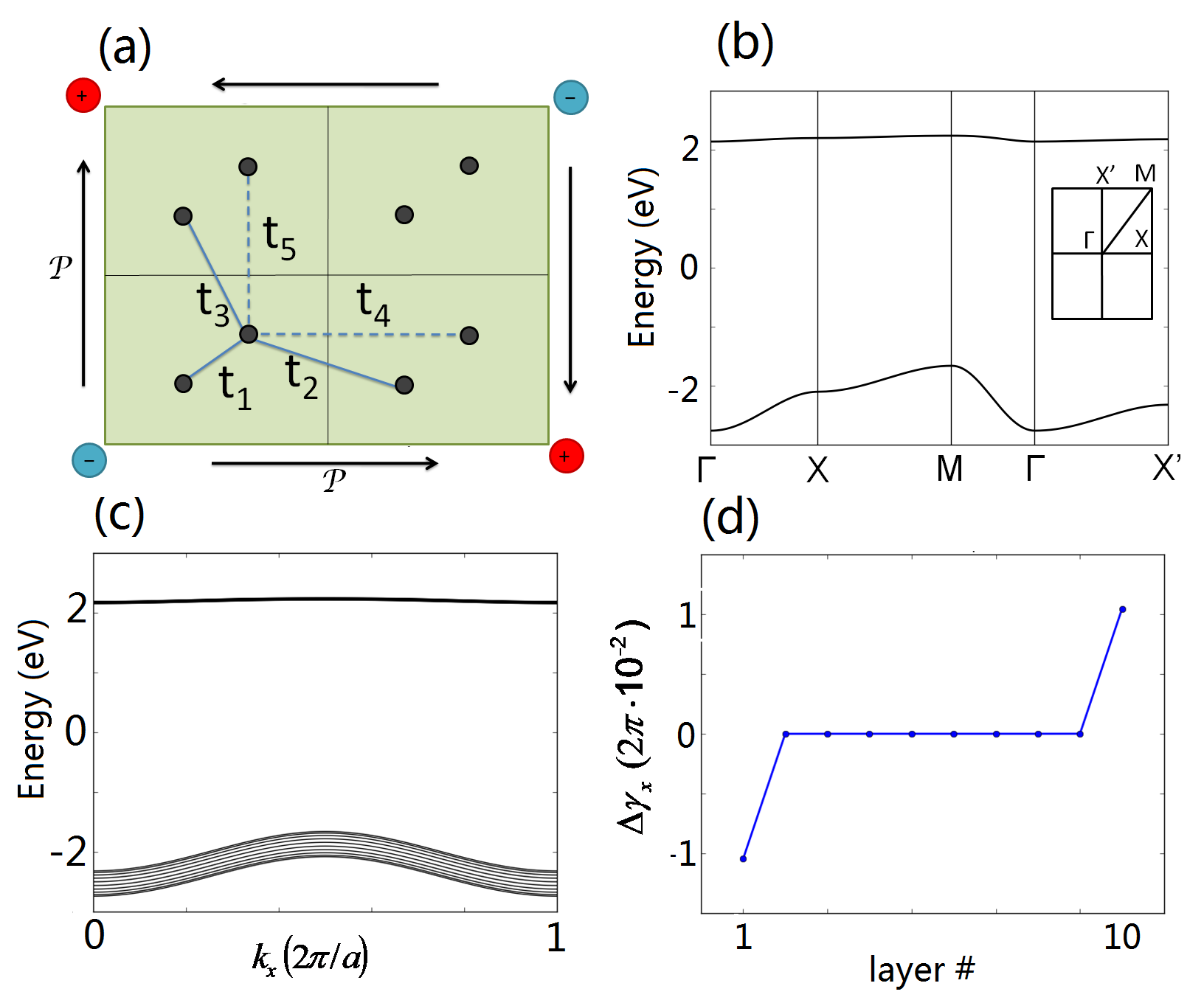}%
 \caption{\label{toymodel}(a) Illustration of the TB model, where four
unit cells are presented.
The atoms are denoted by black dots. Nearest neighbor hoppings $t_1$,
$t_2$ and $t_3$ are shown in solid blue lines. Next-nearest neighbor
hoppings $t_4$ and $t_5$ are shown in dashed blue lines. $\cP$ are shown
by black arrows. The induced $Q_{\textrm {corner}}$ are denoted by red (positive)
and blue (negative) large dots at the corners.
(b) Band structure of the TB model in the $(k_x,k_z)$
 space. The inset shows the high symmetry points in the 2D Brillouin
zone, where $\Gamma$, X, M, $X'$ refer to (0,0), (0,$1\over 2$),
($1\over 2$,$1\over 2$) and ($1\over 2$,0), respectively.
(c) Band structure along $k_x$ for the 2D slab model that is infinite along $x$
while 10-cell-thick in $z$.
(d) Difference between effective $x$ positions of each HWF and that deep
in the bulk.}
 \end{figure}

To illustrate these ideas,
%
%
we start by considering a tight-binding (TB) model of the
simple 2D crystal shown
in Fig.~\ref{toymodel}(a).
We assume a rectangular lattice with an aspect ratio $b/a=0.8$.
There are two atoms
symmetrically located along a diagonal of the rectangular unit cell
with coordinates ($\frac{1}{3},\frac{1}{3}$) and ($\frac{2}{3}$,$\frac{2}{3}$),
so that the bulk crystal has inversion symmetry.  We consider only
one $s$ orbital per atom with onsite energy taken to be zero, and
assume that each atom contributes one electron so that only the lower
band is (doubly) occupied.  We take the nonzero
hoppings to be those shown in Fig.~\ref{toymodel}(a) and
choose their values to be $t_{1}=-2.2$, $t_{2}=-0.15$, and $t_{3}=-0.1$,
$t_{4}=-0.09$ and $t_{5}=-0.06$\, in eV.
The position operators are taken to be diagonal in the
local-orbital representation so that 
$\langle \phi_i|z|\phi_j \rangle = z_i \delta_{ij}$.

We plot the bulk band structure of the TB Hamiltonian
in Fig.~\ref{toymodel}(b). For the selected parameters
the band gap is large compared
to the band widths; in particular, the upper (unoccupied) band is quite flat.
Next we compute the surface polarization of a ribbon cut from
the 2D lattice, taking it to be ten unit cells
thick along $z$ and infinite along $x$.
For the atoms in the surface layers,
the hoppings to the interior atoms are the same as
those described above,
while the hoppings to the vacuum side are set to zero.
We used an equally spaced 60-point
$k_x$ grid. At each $k_x$ the $20\times 20$ Hamiltonian is diagonalized,
resulting in the band structure shown in Fig.~\ref{toymodel}(c).
There are no obvious surface states, and in fact the result
looks almost indistinguishable from a surface projection of the
bulk band structure.
The eigenfunctions $\ket{\psi_n(k_x)}$
are expressed in the tight-binding basis as
$\ket{\psi_n(k_x)}=\sum_j c_{nj}(k_x)\ket{\chi_j({k_x})}$,
where the $\ket{\chi_j({k_x})}$ are the Bloch basis function formed
as a Fourier sum at wavevector $k_x$ of atomic orbitals $\ket{\phi_i}$.
From the ten occupied bands we construct the $10\times 10$
position matrix
$Z_{mn}=\left< \psi_m(k_x)| z | \psi_n(k_x) \right >
=\sum_j z_j c^*_{mj}(k_x)c_{nj}(k_x)$.
Diagonalizing this matrix, we get ten eigenvalues $z_l(k_x)$ that
can each be clearly associated with a particular unit cell layer, and ten
eigenfunctions that are the HWFs.
We label the HWF $\ket{h_{l}(k_x)}$, where $l$ is the layer
index running from the bottom to the top of the ribbon.

Next we calculate $\gamma_{x,l}$, the Berry phase along $k_x$,
for each $l$ using Eq.~(\ref{xpos}).  Deep in the interior
these Berry phases become equal to $\pi$ within numerical precision,
while the Berry phases near the edge are slightly shifted away
from $\pi$, leading to a nonzero surface polarization
as shown in Fig.~\ref{toymodel}(d).

The value of the surface polarization obtained from Eq.~(\ref{surfP}) is
$\cP_x=\pm 2.1\times 10^{-4}\,e$ 
for the top and bottom surfaces respectively.
Similarly we can compute the surface polarizations for the left/right
surfaces using a ribbon ten cells wide in $x$ and infinite along $z$.
We obtain $\cP_z=\pm4.7\times 10^{-4}\,e$ 
along the left and right edges respectively.
At the corners, the
surface polarizations are directed head-to-head
or tail-to-tail, as shown in Fig.~\ref{toymodel}(a).


Given the values of the surface polarizations in the 2D model, we predict
that the charge accumulation at the corner of a finite sample should
equal the sum of the two adjacent surface polarizations, here
$|\cP_x|+|\cP_z|=6.8\times 10^{-4}\,e$.  
To test this, we directly calculate the corner charge in a finite
2D sample, specifically a
$10\times 10$ supercell, large enough to ensure neutrality in the central
region and in the middle of the edges of the sample.
The corner charge is obtained by summing up the on-site charge
differences, relative to the bulk, for atoms in the quadrant
containing the corner.
%
We find $Q_{\textrm{corner}}=6.8\times 10^{-4}\,e$ for the top left and 
bottom right corners, and $-6.8\times 10^{-4}\,e$ for the other two
corners, in agreement with our prediction from the
computed surface polarizations.

 \begin{figure}
 \includegraphics[width=\columnwidth]{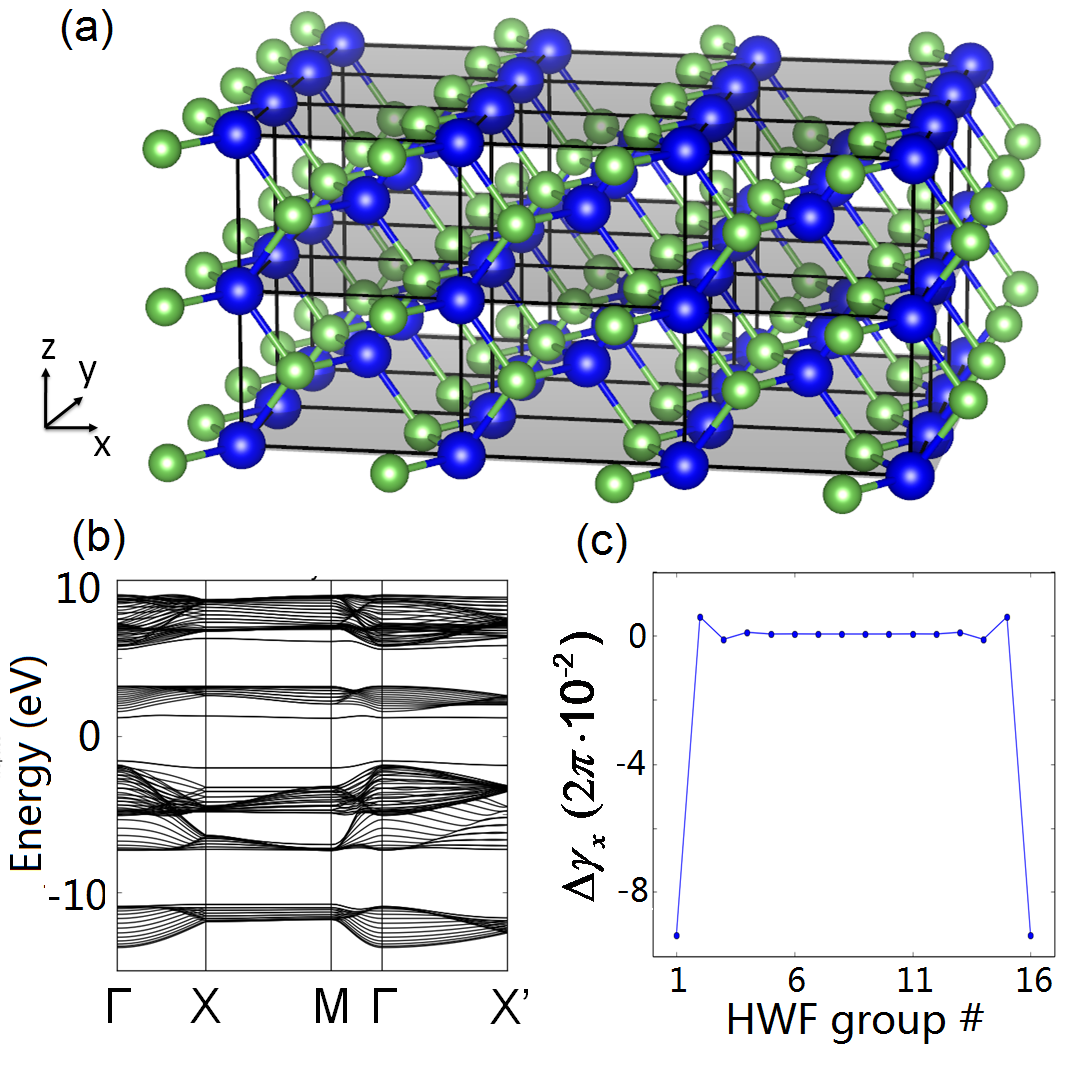}%
 \caption{\label{GaAsslab}(a) Illustration of the GaAs slab studied in the TB model,
 where the blue and green balls
 represent Ga and As atoms, respectively. The grey shaded planes denote
 the (110) family planes. (b) Electronic band
 structure of the GaAs slab in the 2D Brillouin
 zone, with the thickness of 8 cells $z$.
(c) Difference between the $\gamma_x$ of each group of HWFs
and that deep in the bulk.}
 \end{figure}

We now consider a TB model of a generic III-V zincblende
semiconductor, with GaAs as the prototypical example.
The crystal structure is characterized by Ga-As zigzag chains
running along $\left < 110 \right >$.
Although the crystal structure does not have inversion symmetry, the
tetrahedral symmetry forbids a nonzero spontaneous polarization.
We use tight-binding parameters from Ref.~\cite{Cohen-GaAs110}, in which
is shown the bulk bandstructure and density of states.
The unit cell contains two Ga and two As atoms, each with four
$sp^3$ hybridized orbitals and four electrons,
as shown in Fig.~\ref{GaAsslab}(a).
%
%
The position matrix is assumed
to be diagonal and atom-centered in the basis of
tight-binding orbitals.\cite{Bennetto-posMat}

To describe the (110) surface, we construct a slab geometry
as shown in Fig.~\ref{GaAsslab}(a), and we henceforth label
the Cartesian directions as shown there.  That is, the surface,
which is normal to $\hat{z}$, has zigzag chains running along
$\hat{y}$.  Since the two atoms making up these chains are
inequivalent, we expect a surface polarization in the $\hat{x}$
direction.  We take the slab to be eight unit cells thick;
for the atoms in the surface layer, the hoppings to the atoms
inside the slab are the same as in the bulk, while
the hoppings to the vacuum side are set to zero.
At each ($k_x$,$k_y$) of the $100\times 100$  $k$ grid
in the surface BZ, the $128\times 128$
Hamiltonian is diagonalized, and we obtain the band structure
for the slab, shown in Fig.~\ref{GaAsslab}(b).
Surface states are evident as isolated bands.

Next, we diagonalize the 64$\times$64 position matrices 
$Z(k_x,k_y)$ constructed from the eigenstates of the occupied bands.
The eigenvalues, which are the
$z$ coordinates of the HWF centers, can be clearly divided into groups, 
each consisting of four
HWFs $j$ representing the four Ga-As bonds around an As atom,
each group being associated with one of the 16 atomic layers $l$.
In this case, it is more useful to calculate the Berry
phase of each group of HWFs
rather than of each single HWF~\cite{polar-VKS-1993}.

As expected, the Berry phase in the $\hat{y}$ direction along
the zigzag chain is found to be zero, but in the $\hat{x}$ direction
it is nonzero for the HWF groups near the top and bottom surfaces
of the slab.  Thus, we confirm that there is a nonzero
surface polarization $\cP_x$.
We plot the difference between the Berry phase $\gamma_x$ of each group of HWFs
and that for the bulk in Fig.~\ref{GaAsslab}(c).
By summing up the contributions from each group of HWFs from the center
of the bulk to one surface, the total surface polarization is
found to be 0.178 
$e/L$. Here 
%
%
$L=a/\sqrt{2}$ is the repeat length of the zigzag chain, i.e.,
the surface cell dimension along $\hat{y}$, where $a$ is the
surface lattice constant along $\hat{x}$.
%
%
Subdividing the dominant surface-group contribution further,
we find that the surface polarization comes mainly from the
surface-most HWF, corresponding to a shift of the center of the
dangling bond on the surface As atom.
%
%

\begin{figure}
 \includegraphics[width=\columnwidth]{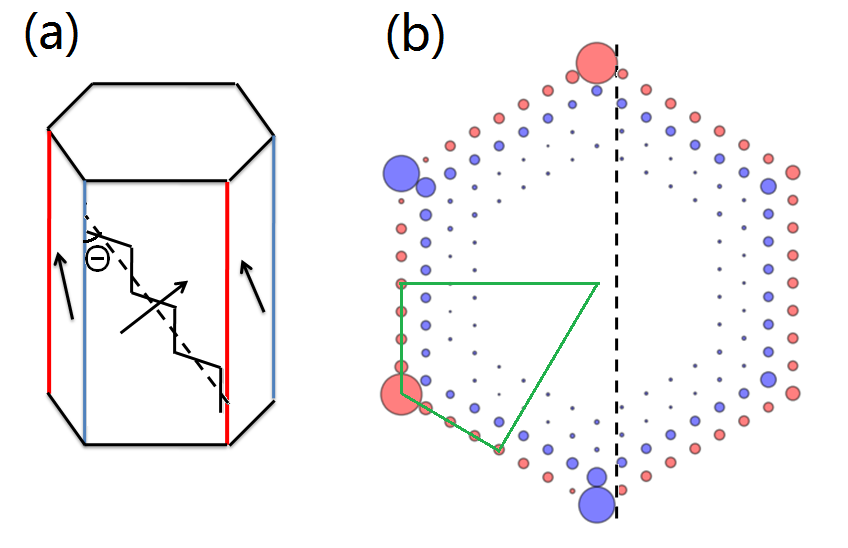}%
 \caption{\label{GaAswire}(a) Overhead view of the hexagonal
 GaAs nanowire. 
The dashed black line, which meets the edges along [111] at an angle
of $\theta=35.26^\circ$, shows the direction along which
the zigzag surface chains run.
The relevant surface polarizations at the side
surfaces are denoted by
 black arrows. The blue and red vertical edges mean net negative
 and positive edge charge distributions, respectively.
(b) On-site charge distribution summed over the trilayer.
Red and blue dots
 represent positive and negative net charges, respectively. The
 sizes of dots indicate the magnitudes of the on-site net charge.
the left and right regions to the dashed vertical line show the total and symmetric
part of onsite charges, respectively.}
 \end{figure}

The surface polarization on the \{110\} surfaces predicts
an accumulated line charge for the common edge of two such surfaces.
In order to demonstrate this effect, we consider a hexagonal wire of GaAs
that is infinite along [111], with
a periodicity corresponding to three of the GaAs
buckled (111) layers.
In this case,
the six side surfaces of the wire are all \{110\} planes: ($1\bar
10$), ($10\bar 1$), ($0 1\bar 1$), ($\bar 1 10$), ($\bar 101$),
and ($0\bar1 1$). 
As shown in Fig.~\ref{GaAswire}(a),
on each side facet the surface polarization is perpendicular to
the zigzag chains, forming a pattern of $\bcP$ vectors shown as
black arrows.
The surface polarizations for each neighboring pair of
side facets have a common component along [111],
but are head-to-head or tail-to-tail for the component
normal to [111], leading to alternating positive
and negative line charges for the six edges as shown.
According to Eq.~(\ref{eq:edgecharge}), we expect the line charge per
three-layer vertical period
to be $Q_{3L}=2\cP\cos\theta\cdot 3L\cos\theta=0.71 e$, 
where the $3L\cos\theta$ factor is the vertical period.

For comparison, we directly calculate the edge charges per trilayer period
in a nanowire with a radius of 8 atoms.
%
%
We sum up the site populations within the TB model with a 60-point $k$ grid along [111].
The onsite charge is the difference from the bulk value.
The computed onsite charges are shown in the left half
of Fig.~\ref{GaAswire}(b), while the right half shows the
corresponding results after averaging with a 60$^\circ$-rotated
version of itself.  The surface charges decay rapidly into the
bulk, leading to a neutral bulk state inside the nanowire.  Also,
a surface dipole density normal to the surface is clearly visible,
especially in the orientationally averaged results.  However, we
are interested in the accumulation at the edges, which is obviously
present in the unaveraged results in the left half of the figure.
The edge charge is calculated by summing up the onsite charges in
the wedge-shaped region illustrated in Fig. \ref{GaAswire}(b),
using a weight of 1/2 for atoms located on its radial edges.
The total edge charge per trilayer is found to be $\pm$0.71$e$,
in
agreement with the value predicted using the
previously calculated surface polarization.



We emphasize that this numerical value is not
intended to be realistic for GaAs.  A more accurate estimate would
require the use of an improved tight-binding model
and treatment of surface relaxations and dielectric screening
effects, or better, direct first-principles calculations.
%
%
Our purpose here has been to show
that the surface
polarization as defined here correctly predicts edge charges.
We note that an analysis based on maximally localized Wannier
functions~\cite{RMP-WF} is also possible.
However, we believe our HWF-based
approach is more natural, as the Wannier transformation
is only done in the needed direction and no iterative construction
is required.

We stress that the concept of surface polarization $\bcP$
is quite general, occurring whenever the surface symmetry is
low enough.  In some cases this can arise from
a spontaneous symmetry-lowering surface relaxation
or reconstruction, allowing ``surface ferroelectricity'' if
it is switchable.  In other cases, as for GaAs (110), the ideal surface
space group already has low enough symmetry to allow a nonzero
$\bcP$. This will occur quite generally for low-angle vicinal
surfaces. The concept also applies to planar defects such as
domain walls, stacking faults, and twin boundaries, and to
heterointerfaces; if $\bcP$ is present within this plane, it may induce
a line charge where the plane intersects the surface.  Such edge and
line charges are potentially observable using electric force
microscopy~\cite{valdre2006electric}
, electron holography~\cite{wolf-apl11}, or other experimental methods.
Finally we note that the concept of surface polarization may become
  more subtle in the presence of orbital magnetization, which we
  have omitted from our considerations here.
%


In summary, we have formulated the concept of surface polarization,
i.e., the dipole moment per unit area {\it parallel} to the surface,
which can exist whenever the surface symmetry is low enough.
Using TB models we have computed the surface polarizations for
a 2D toy model and a generic III-V zincblende semiconductor,
and shown that the predicted corner or edge charges are in good
agreement with direct calculations.
We point out that surface and interface polarizations can be responsible
for observable effects, and perhaps even desirable functionality,
in a broad range of insulating materials systems.

We acknowledge helpful discussions with Hongbin
Zhang, Jianpeng Liu and Massimiliano Stengel. This work is supported by NSF DMR-14-08838 and ONR N00014-11-1-0665

\bibliography{surfaceP}

\begin{thebibliography}{17}%
\makeatletter
\providecommand \@ifxundefined [1]{%
 \@ifx{#1\undefined}
}%
\providecommand \@ifnum [1]{%
 \ifnum #1\expandafter \@firstoftwo
 \else \expandafter \@secondoftwo
 \fi
}%
\providecommand \@ifx [1]{%
 \ifx #1\expandafter \@firstoftwo
 \else \expandafter \@secondoftwo
 \fi
}%
\providecommand \natexlab [1]{#1}%
\providecommand \enquote  [1]{``#1''}%
\providecommand \bibnamefont  [1]{#1}%
\providecommand \bibfnamefont [1]{#1}%
\providecommand \citenamefont [1]{#1}%
\providecommand \href@noop [0]{\@secondoftwo}%
\providecommand \href [0]{\begingroup \@sanitize@url \@href}%
\providecommand \@href[1]{\@@startlink{#1}\@@href}%
\providecommand \@@href[1]{\endgroup#1\@@endlink}%
\providecommand \@sanitize@url [0]{\catcode `\\12\catcode `\$12\catcode
  `\&12\catcode `\#12\catcode `\^12\catcode `\_12\catcode `\%12\relax}%
\providecommand \@@startlink[1]{}%
\providecommand \@@endlink[0]{}%
\providecommand \url  [0]{\begingroup\@sanitize@url \@url }%
\providecommand \@url [1]{\endgroup\@href {#1}{\urlprefix }}%
\providecommand \urlprefix  [0]{URL }%
\providecommand \Eprint [0]{\href }%
\providecommand \doibase [0]{http://dx.doi.org/}%
\providecommand \selectlanguage [0]{\@gobble}%
\providecommand \bibinfo  [0]{\@secondoftwo}%
\providecommand \bibfield  [0]{\@secondoftwo}%
\providecommand \translation [1]{[#1]}%
\providecommand \BibitemOpen [0]{}%
\providecommand \bibitemStop [0]{}%
\providecommand \bibitemNoStop [0]{.\EOS\space}%
\providecommand \EOS [0]{\spacefactor3000\relax}%
\providecommand \BibitemShut  [1]{\csname bibitem#1\endcsname}%
\let\auto@bib@innerbib\@empty
\bibitem [{\citenamefont {King-Smith}\ and\ \citenamefont
  {Vanderbilt}(1993)}]{polar-KSV-1993}%
  \BibitemOpen
  \bibfield  {author} {\bibinfo {author} {\bibfnamefont {R.~D.}\ \bibnamefont
  {King-Smith}}\ and\ \bibinfo {author} {\bibfnamefont {D.}~\bibnamefont
  {Vanderbilt}},\ }\href {\doibase 10.1103/PhysRevB.47.1651} {\bibfield
  {journal} {\bibinfo  {journal} {Phys. Rev. B}\ }\textbf {\bibinfo {volume}
  {47}},\ \bibinfo {pages} {1651} (\bibinfo {year} {1993})}\BibitemShut
  {NoStop}%
\bibitem [{\citenamefont {Resta}(1994)}]{polar-RMP-1994}%
  \BibitemOpen
  \bibfield  {author} {\bibinfo {author} {\bibfnamefont {R.}~\bibnamefont
  {Resta}},\ }\href {\doibase 10.1103/RevModPhys.66.899} {\bibfield  {journal}
  {\bibinfo  {journal} {Rev. Mod. Phys.}\ }\textbf {\bibinfo {volume} {66}},\
  \bibinfo {pages} {899} (\bibinfo {year} {1994})}\BibitemShut {NoStop}%
\bibitem [{\citenamefont {Vanderbilt}\ and\ \citenamefont
  {King-Smith}(1993)}]{polar-VKS-1993}%
  \BibitemOpen
  \bibfield  {author} {\bibinfo {author} {\bibfnamefont {D.}~\bibnamefont
  {Vanderbilt}}\ and\ \bibinfo {author} {\bibfnamefont {R.~D.}\ \bibnamefont
  {King-Smith}},\ }\href@noop {} {\bibfield  {journal} {\bibinfo  {journal}
  {Phys. Rev. B}\ }\textbf {\bibinfo {volume} {48}},\ \bibinfo {pages} {4442}
  (\bibinfo {year} {1993})}\BibitemShut {NoStop}%
\bibitem [{\citenamefont {B\'anyai}\ \emph {et~al.}(1992)\citenamefont
  {B\'anyai}, \citenamefont {Gilliot}, \citenamefont {Hu},\ and\ \citenamefont
  {Koch}}]{surfP-prb}%
  \BibitemOpen
  \bibfield  {author} {\bibinfo {author} {\bibfnamefont {L.}~\bibnamefont
  {B\'anyai}}, \bibinfo {author} {\bibfnamefont {P.}~\bibnamefont {Gilliot}},
  \bibinfo {author} {\bibfnamefont {Y.~Z.}\ \bibnamefont {Hu}}, \ and\ \bibinfo
  {author} {\bibfnamefont {S.~W.}\ \bibnamefont {Koch}},\ }\href {\doibase
  10.1103/PhysRevB.45.14136} {\bibfield  {journal} {\bibinfo  {journal} {Phys.
  Rev. B}\ }\textbf {\bibinfo {volume} {45}},\ \bibinfo {pages} {14136}
  (\bibinfo {year} {1992})}\BibitemShut {NoStop}%
\bibitem [{\citenamefont {Wen}\ \emph {et~al.}(2003)\citenamefont {Wen},
  \citenamefont {Huang}, \citenamefont {Yang}, \citenamefont {Lu},\ and\
  \citenamefont {Sheng}}]{surfacePNMat}%
  \BibitemOpen
  \bibfield  {author} {\bibinfo {author} {\bibfnamefont {W.}~\bibnamefont
  {Wen}}, \bibinfo {author} {\bibfnamefont {X.}~\bibnamefont {Huang}}, \bibinfo
  {author} {\bibfnamefont {S.}~\bibnamefont {Yang}}, \bibinfo {author}
  {\bibfnamefont {K.}~\bibnamefont {Lu}}, \ and\ \bibinfo {author}
  {\bibfnamefont {P.}~\bibnamefont {Sheng}},\ }\href {\doibase 10.1038/nmat993}
  {\bibfield  {journal} {\bibinfo  {journal} {Nat. Mater.}\ }\textbf {\bibinfo
  {volume} {2}},\ \bibinfo {pages} {727} (\bibinfo {year} {2003})}\BibitemShut
  {NoStop}%
\bibitem [{\citenamefont {Okada}\ \emph {et~al.}(2013)\citenamefont {Okada},
  \citenamefont {Serbyn}, \citenamefont {Lin}, \citenamefont {Walkup},
  \citenamefont {Zhou}, \citenamefont {Dhital}, \citenamefont {Neupane},
  \citenamefont {Xu}, \citenamefont {Wang}, \citenamefont {Sankar},
  \citenamefont {Chou}, \citenamefont {Bansil}, \citenamefont {Hasan},
  \citenamefont {Wilson}, \citenamefont {Fu},\ and\ \citenamefont
  {Madhavan}}]{Science-TCI-surfaces}%
  \BibitemOpen
  \bibfield  {author} {\bibinfo {author} {\bibfnamefont {Y.}~\bibnamefont
  {Okada}}, \bibinfo {author} {\bibfnamefont {M.}~\bibnamefont {Serbyn}},
  \bibinfo {author} {\bibfnamefont {H.}~\bibnamefont {Lin}}, \bibinfo {author}
  {\bibfnamefont {D.}~\bibnamefont {Walkup}}, \bibinfo {author} {\bibfnamefont
  {W.}~\bibnamefont {Zhou}}, \bibinfo {author} {\bibfnamefont {C.}~\bibnamefont
  {Dhital}}, \bibinfo {author} {\bibfnamefont {M.}~\bibnamefont {Neupane}},
  \bibinfo {author} {\bibfnamefont {S.}~\bibnamefont {Xu}}, \bibinfo {author}
  {\bibfnamefont {Y.~J.}\ \bibnamefont {Wang}}, \bibinfo {author}
  {\bibfnamefont {R.}~\bibnamefont {Sankar}}, \bibinfo {author} {\bibfnamefont
  {F.}~\bibnamefont {Chou}}, \bibinfo {author} {\bibfnamefont {A.}~\bibnamefont
  {Bansil}}, \bibinfo {author} {\bibfnamefont {M.~Z.}\ \bibnamefont {Hasan}},
  \bibinfo {author} {\bibfnamefont {S.~D.}\ \bibnamefont {Wilson}}, \bibinfo
  {author} {\bibfnamefont {L.}~\bibnamefont {Fu}}, \ and\ \bibinfo {author}
  {\bibfnamefont {V.}~\bibnamefont {Madhavan}},\ }\href {\doibase
  10.1126/science.1239451} {\bibfield  {journal} {\bibinfo  {journal}
  {Science}\ }\textbf {\bibinfo {volume} {341}},\ \bibinfo {pages} {1496}
  (\bibinfo {year} {2013})}\BibitemShut {NoStop}%
\bibitem [{\citenamefont {Gupta}\ \emph {et~al.}(2013)\citenamefont {Gupta},
  \citenamefont {Mahadevan}, \citenamefont {Mavropoulos},\ and\ \citenamefont
  {Le\ifmmode \check{z}\else \v{z}\fi{}ai\ifmmode~\acute{c}\else
  \'{c}\fi{}}}]{SCOferro}%
  \BibitemOpen
  \bibfield  {author} {\bibinfo {author} {\bibfnamefont {K.}~\bibnamefont
  {Gupta}}, \bibinfo {author} {\bibfnamefont {P.}~\bibnamefont {Mahadevan}},
  \bibinfo {author} {\bibfnamefont {P.}~\bibnamefont {Mavropoulos}}, \ and\
  \bibinfo {author} {\bibfnamefont {M.}~\bibnamefont {Le\ifmmode \check{z}\else
  \v{z}\fi{}ai\ifmmode~\acute{c}\else \'{c}\fi{}}},\ }\href {\doibase
  10.1103/PhysRevLett.111.077601} {\bibfield  {journal} {\bibinfo  {journal}
  {Phys. Rev. Lett.}\ }\textbf {\bibinfo {volume} {111}},\ \bibinfo {pages}
  {077601} (\bibinfo {year} {2013})}\BibitemShut {NoStop}%
\bibitem [{\citenamefont {Sgiarovello}\ \emph {et~al.}(2001)\citenamefont
  {Sgiarovello}, \citenamefont {Peressi},\ and\ \citenamefont
  {Resta}}]{Resta-HWF}%
  \BibitemOpen
  \bibfield  {author} {\bibinfo {author} {\bibfnamefont {C.}~\bibnamefont
  {Sgiarovello}}, \bibinfo {author} {\bibfnamefont {M.}~\bibnamefont
  {Peressi}}, \ and\ \bibinfo {author} {\bibfnamefont {R.}~\bibnamefont
  {Resta}},\ }\href {\doibase 10.1103/PhysRevB.64.115202} {\bibfield  {journal}
  {\bibinfo  {journal} {Phys. Rev. B}\ }\textbf {\bibinfo {volume} {64}},\
  \bibinfo {pages} {115202} (\bibinfo {year} {2001})}\BibitemShut {NoStop}%
\bibitem [{\citenamefont {Wu}\ \emph {et~al.}(2006)\citenamefont {Wu},
  \citenamefont {Di\'eguez}, \citenamefont {Rabe},\ and\ \citenamefont
  {Vanderbilt}}]{wu-prl06}%
  \BibitemOpen
  \bibfield  {author} {\bibinfo {author} {\bibfnamefont {X.}~\bibnamefont
  {Wu}}, \bibinfo {author} {\bibfnamefont {O.}~\bibnamefont {Di\'eguez}},
  \bibinfo {author} {\bibfnamefont {K.~M.}\ \bibnamefont {Rabe}}, \ and\
  \bibinfo {author} {\bibfnamefont {D.}~\bibnamefont {Vanderbilt}},\ }\href
  {\doibase 10.1103/PhysRevLett.97.107602} {\bibfield  {journal} {\bibinfo
  {journal} {Phys. Rev. Lett.}\ }\textbf {\bibinfo {volume} {97}},\ \bibinfo
  {pages} {107602} (\bibinfo {year} {2006})}\BibitemShut {NoStop}%
\bibitem [{\citenamefont {Soluyanov}\ and\ \citenamefont
  {Vanderbilt}(2011)}]{soluyanov-prb11b}%
  \BibitemOpen
  \bibfield  {author} {\bibinfo {author} {\bibfnamefont {A.~A.}\ \bibnamefont
  {Soluyanov}}\ and\ \bibinfo {author} {\bibfnamefont {D.}~\bibnamefont
  {Vanderbilt}},\ }\href {\doibase {10.1103/PhysRevB.83.235401}} {\bibfield
  {journal} {\bibinfo  {journal} {{Phys. Rev. B}}\ }\textbf {\bibinfo {volume}
  {{83}}},\ \bibinfo {pages} {{235401}} (\bibinfo {year} {{2011}})}\BibitemShut
  {NoStop}%
\bibitem [{\citenamefont {Taherinejad}\ \emph {et~al.}(2014)\citenamefont
  {Taherinejad}, \citenamefont {Garrity},\ and\ \citenamefont
  {Vanderbilt}}]{taherinejad-prb13}%
  \BibitemOpen
  \bibfield  {author} {\bibinfo {author} {\bibfnamefont {M.}~\bibnamefont
  {Taherinejad}}, \bibinfo {author} {\bibfnamefont {K.~F.}\ \bibnamefont
  {Garrity}}, \ and\ \bibinfo {author} {\bibfnamefont {D.}~\bibnamefont
  {Vanderbilt}},\ }\href {\doibase 10.1103/PhysRevB.89.115102} {\bibfield
  {journal} {\bibinfo  {journal} {Phys. Rev. B}\ }\textbf {\bibinfo {volume}
  {89}},\ \bibinfo {pages} {115102} (\bibinfo {year} {2014})}\BibitemShut
  {NoStop}%
\bibitem [{\citenamefont {Marzari}\ and\ \citenamefont
  {Vanderbilt}(1997)}]{Marzari-MLWF}%
  \BibitemOpen
  \bibfield  {author} {\bibinfo {author} {\bibfnamefont {N.}~\bibnamefont
  {Marzari}}\ and\ \bibinfo {author} {\bibfnamefont {D.}~\bibnamefont
  {Vanderbilt}},\ }\href {\doibase 10.1103/PhysRevB.56.12847} {\bibfield
  {journal} {\bibinfo  {journal} {Phys. Rev. B}\ }\textbf {\bibinfo {volume}
  {56}},\ \bibinfo {pages} {12847} (\bibinfo {year} {1997})}\BibitemShut
  {NoStop}%
\bibitem [{\citenamefont {Joannopoulos}\ and\ \citenamefont
  {Cohen}(1974)}]{Cohen-GaAs110}%
  \BibitemOpen
  \bibfield  {author} {\bibinfo {author} {\bibfnamefont {J.~D.}\ \bibnamefont
  {Joannopoulos}}\ and\ \bibinfo {author} {\bibfnamefont {M.~L.}\ \bibnamefont
  {Cohen}},\ }\href {\doibase 10.1103/PhysRevB.10.5075} {\bibfield  {journal}
  {\bibinfo  {journal} {Phys. Rev. B}\ }\textbf {\bibinfo {volume} {10}},\
  \bibinfo {pages} {5075} (\bibinfo {year} {1974})}\BibitemShut {NoStop}%
\bibitem [{\citenamefont {Bennetto}\ and\ \citenamefont
  {Vanderbilt}(1996)}]{Bennetto-posMat}%
  \BibitemOpen
  \bibfield  {author} {\bibinfo {author} {\bibfnamefont {J.}~\bibnamefont
  {Bennetto}}\ and\ \bibinfo {author} {\bibfnamefont {D.}~\bibnamefont
  {Vanderbilt}},\ }\href {\doibase 10.1103/PhysRevB.53.15417} {\bibfield
  {journal} {\bibinfo  {journal} {Phys. Rev. B}\ }\textbf {\bibinfo {volume}
  {53}},\ \bibinfo {pages} {15417} (\bibinfo {year} {1996})}\BibitemShut
  {NoStop}%
\bibitem [{\citenamefont {Marzari}\ \emph {et~al.}(2012)\citenamefont
  {Marzari}, \citenamefont {Mostofi}, \citenamefont {Yates}, \citenamefont
  {Souza},\ and\ \citenamefont {Vanderbilt}}]{RMP-WF}%
  \BibitemOpen
  \bibfield  {author} {\bibinfo {author} {\bibfnamefont {N.}~\bibnamefont
  {Marzari}}, \bibinfo {author} {\bibfnamefont {A.~A.}\ \bibnamefont
  {Mostofi}}, \bibinfo {author} {\bibfnamefont {J.~R.}\ \bibnamefont {Yates}},
  \bibinfo {author} {\bibfnamefont {I.}~\bibnamefont {Souza}}, \ and\ \bibinfo
  {author} {\bibfnamefont {D.}~\bibnamefont {Vanderbilt}},\ }\href {\doibase
  10.1103/RevModPhys.84.1419} {\bibfield  {journal} {\bibinfo  {journal} {Rev.
  Mod. Phys.}\ }\textbf {\bibinfo {volume} {84}},\ \bibinfo {pages} {1419}
  (\bibinfo {year} {2012})}\BibitemShut {NoStop}%
\bibitem [{\citenamefont {Valdr{\`e}}(2006)}]{valdre2006electric}%
  \BibitemOpen
  \bibfield  {author} {\bibinfo {author} {\bibfnamefont {G.}~\bibnamefont
  {Valdr{\`e}}},\ }\href@noop {} {\bibfield  {journal} {\bibinfo  {journal}
  {Imaging \& Microscopy}\ }\textbf {\bibinfo {volume} {8}},\ \bibinfo {pages}
  {44} (\bibinfo {year} {2006})}\BibitemShut {NoStop}%
\bibitem [{\citenamefont {Wolf}\ \emph {et~al.}(2011)\citenamefont {Wolf},
  \citenamefont {Lichte}, \citenamefont {Pozzi}, \citenamefont {Prete},\ and\
  \citenamefont {Lovergine}}]{wolf-apl11}%
  \BibitemOpen
  \bibfield  {author} {\bibinfo {author} {\bibfnamefont {D.}~\bibnamefont
  {Wolf}}, \bibinfo {author} {\bibfnamefont {H.}~\bibnamefont {Lichte}},
  \bibinfo {author} {\bibfnamefont {G.}~\bibnamefont {Pozzi}}, \bibinfo
  {author} {\bibfnamefont {P.}~\bibnamefont {Prete}}, \ and\ \bibinfo {author}
  {\bibfnamefont {N.}~\bibnamefont {Lovergine}},\ }\href@noop {} {\bibfield
  {journal} {\bibinfo  {journal} {Applied Physics Letters}\ }\textbf {\bibinfo
  {volume} {98}},\ \bibinfo {pages} {264103} (\bibinfo {year}
  {2011})}\BibitemShut {NoStop}%
\end{thebibliography}%
\end{document}